\documentstyle[12pt]{article}

\newcommand{\qed}{\hbox{\rule{6pt}{9pt}}}

\catcode `\@=11
\def\@cite#1#2{#1\if@tempswa , #2\fi}
\catcode `\@=12
\newcommand{\upcite}[1]{${}^{\mbox{\scriptsize \cite{#1})}}$}

\newcommand{\uprcite}[2]{${}^{\mbox{\scriptsize \cite{#1} - \cite{#2})}}$}

\begin{document}
\vspace*{1mm}

\begin{center}
{\large\bf An Upper Bound for the Spin-Wave Spectrum} \\
\medskip

{\large\bf of the Heisenberg Antiferromagnet\footnote{To be published
in J. Phys. Soc. Jpn. {\bf 63} (1994) No. 7.}}
\bigskip
\bigskip
\smallskip

{\large Tsutomu Momoi}
\bigskip

\bigskip

Department of Physics, University of Tokyo,
Hongo 7-3-1, Bunkyo-ku, Tokyo 113, Japan
\bigskip

\end{center}
\medskip

\begin{abstract}
\begin{normalsize}
We study the spin-wave spectrum of the Heisenberg antiferromagnet on
a bipartite lattice.
The spin-wave spectrum on a N\'eel-ordered ground state is bounded as
$\varepsilon (\mbox{\boldmath $k$})\le c\vert \mbox{\boldmath $k$}\vert$, where
$c$ gives an upper bound for the spin-wave velocity. In the large-$S$ limit,
the upper bound $c$ coincides with the result of the spin-wave theory.
\end{normalsize}
\end{abstract}
\bigskip

\smallskip

\bigskip

\bigskip

Ground-state properties of the quantum Heisenberg antiferromagnet
have been studied extensively.
The existence of N\'eel order in the ground state has been proven
for the three-dimensional spin-$S$ Heisenberg
antiferromagnets on bipartite lattices.\uprcite{Dyson}{Koma1993A}
In the two-dimensional systems, the occurrence of symmetry breaking at
$T=0$ has
been proven for higher spin ($S\ge 1$) models and for anisotropic
models.\uprcite{Koma1993A}{Nishimori1989}
In the spin $1/2$ Heisenberg antiferromagnet, many theoretical and numerical
studies indicate that the system has a long-range order, though there is
no rigorous proof (to date).

The Nambu-Goldstone theorem for quantum spin
systems\upcite{Wreszinski} states that, if there is a N\'eel order,
there are gapless elementary excitations.
In the Heisenberg antiferromagnet, the spin-wave theory\upcite{Anderson,RKubo}
succeeded in giving a precise description of the Goldstone bosons, i.e.,
magnons. The linear spin-wave
theory predicts the gapless $k$-linear dispersion relation,
$\varepsilon (\mbox{\boldmath $k$}) \simeq 2\sqrt{d}SJ\vert \mbox{\boldmath
$k$}
\vert$, for small $k$.

In this letter
we consider the quantum Heisenberg and $XXZ$
antiferromagnets on the
$d$-dimensional $L\times \cdots \times L$ hypercubic lattice
$\Lambda \subset {\rm Z}^d$ and we consider the case of $d\ge2$.
The Hamiltonian is given by
\begin{equation}\label{Hamiltonian}
{\cal H}_\Lambda=J\sum_{\langle i,j\rangle \in\Lambda}
(S_i^x S_j^x+S_i^y S_j^y+\Delta S_i^z S_j^z),
\end{equation}
where $0\le \Delta \le 1$ and the summation runs over all the
nearest-neighbor sites.
The system size of $\Lambda$ is $N(=L^d)$.
We show an upper bound for the spectrum of
elementary excitations on the N\'eel-ordered ground states.
We find that the excitation spectrum is bounded as
$\varepsilon (\mbox{\boldmath $k$})\le c\vert \mbox{\boldmath $k$}\vert$,
where the constant $c$ gives an upper bound for the spin-wave velocity.
For $S \gg 1$, the upper bound $c$ coincides with the result of the spin-wave
approximation. In the Appendix,
we show some rigorous results for the large-$S$
Heisenberg antiferromagnet.

Let us consider the system under the small staggered magnetic
field whose Hamiltonian is
\begin{equation}\label{Hamiltonian_XXZ}
{\cal H}_\Lambda (B) = {\cal H}_\Lambda -B {\cal M}_\Lambda,
\end{equation}
where
\begin{equation}
{\cal M}_\Lambda=\sum_{i\in A}S^x_i - \sum_{i\in B}S^x_i.
\end{equation}
We denote the normalized ground state as
$\vert \Phi_{{\rm GS},B} \rangle $.
We take a thermodynamic limit applying an infinitesimally small field $B$.
The staggered magnetization is given by
\begin{equation}
m_{\rm s} = \lim_{B\downarrow 0}\lim_{\Lambda \uparrow \infty}
{1 \over N} \langle \Phi_{{\rm GS},B} \vert {\cal M}_\Lambda
\vert \Phi_{{\rm GS},B} \rangle.
\end{equation}
As an excited state, we consider the following standard trial state
\begin{equation}\label{st:Trial}
\vert \Psi_{B} (\mbox{\boldmath $k$}) \rangle =
S^z_{k} \vert \Phi_{{\rm GS},B} \rangle
/\Vert S^z_{k} \vert \Phi_{{\rm GS},B} \rangle \Vert,
\end{equation}
where\ \ $S^z_{k}=N^{-1/2} \sum_i S^z_i
\exp (i\mbox{\boldmath $k$}\cdot \mbox{\boldmath $r$}_i )$,
\ \ $\mbox{\boldmath $k$}=(k_1,k_2,\ldots,k_d)$\ \ and \\
$\Vert S^z_k \vert \Phi_{{\rm GS},B}\rangle \Vert =
\langle \Phi_{{\rm GS},B} \vert S^z_{-k} S^z_k
\vert \Phi_{{\rm GS},B}\rangle^{1/2} $.
When spins lie in the $xy$ plane, the operation of $S^z_i$ flips
the spin at the site $i$.
The excitation energy of $\vert \Psi_{B} (\mbox{\boldmath $k$}) \rangle $
is given by
\begin{eqnarray}
\lefteqn{\varepsilon(\mbox{\boldmath $k$}) =
\lim_{B\downarrow 0}\lim_{\Lambda \uparrow \infty}
\{ \langle \Psi_{B} (\mbox{\boldmath $k$}) \vert {\cal H}_\Lambda (B) \vert
\Psi_{B} (\mbox{\boldmath $k$})\rangle }\\
& &\mbox{\hspace{2cm}}-\langle \Phi_{{\rm GS},B} \vert {\cal H}_\Lambda (B)
\vert \Phi_{{\rm GS},B} \rangle \}.\nonumber
\end{eqnarray}
We take the momentum $\mbox{\boldmath $k$}$
as $\mbox{\boldmath $k$} \ne 0$,
$\mbox{\boldmath $k$} \ne (\pi,\ldots,\pi)$
and $k_m =2\pi l_m / L$
$(0\le l_m \le L-1)$, so that
$\vert \Phi_{{\rm GS},B} \rangle$ and
$\vert \Psi_{B} (\mbox{\boldmath $k$}) \rangle$ are
orthogonal. This trial state is called the Bijl-Feynman single-mode
approximation.\upcite{Bijl,Feynman} A similar trial state
was studied by Horsch and von der
Linden in the two-dimensional system with no magnetic field.\upcite{Horsch}
Using the state (\ref{st:Trial}), we show an upper bound
of the spin-wave spectrum as follows.\footnote{After the submission of this
manuscript, we learned that Stringari had previously obtained a better upper
bound of the spin-wave spectrum.\upcite{Stringari}
We thank Mr.~K.~Totsuka for letting us
know Stringari's paper (ref.~\cite{Stringari}).
New points of the present paper are that we obtain both upper and
lower bounds of the excitation spectrum of the Feynman state
(\ref{st:Trial}) and that we give rigorous arguments on the large-$S$ limit. }
\bigskip

\bigskip

{\bf Theorem:}
If the ground state has a N\'eel order, i.e., if $m_{\rm s}>0$,
the energy spectrum $\varepsilon (\mbox{\boldmath $k$})$ is bounded as
\begin{equation}\label{ineq:Th3}
\varepsilon (\mbox{\boldmath $k$}) \le {2d J (\rho_x +\rho_y )\over {m_{\rm
s}}^2 }
\sqrt{\rho_x (1+\Delta \gamma_k) + \rho_z (\Delta + \gamma_k) }
\sqrt{1-{\gamma_k}},
\end{equation}
where
\begin{equation}
\rho_\alpha = -\lim_{B\downarrow 0}\lim_{\Lambda \uparrow \infty}
\frac{1}{Nd}\langle \Phi_{{\rm GS},B} \vert
\sum_{\langle i,j\rangle\in \Lambda} S^\alpha_i S^\alpha_j
\vert \Phi_{{\rm GS},B} \rangle
\end{equation}
for $\alpha =x,$ $y$ and $z$, and
\begin{equation}
\gamma_k = {1\over d}\sum_{i=1}^{d} \cos k_i .
\end{equation}
\bigskip

\bigskip

{\it Proof.} As in ref. \cite{Horsch}, the excitation energy of
$\vert \Psi_{B} (\mbox{\boldmath $k$}) \rangle$ is calculated as
\begin{eqnarray}
\lefteqn{\langle \Psi_{B} (\mbox{\boldmath $k$})\vert {\cal H}_\Lambda (B)
\vert \Psi_{B} (\mbox{\boldmath $k$}) \rangle
-\langle \Phi_{{\rm GS},B} \vert {\cal H}_\Lambda (B)
\vert \Phi_{{\rm GS},B} \rangle
=
\frac{ \langle \Phi_{{\rm GS},B} \vert
[[ S^z_{-k},{\cal H}_\Lambda (B)], S^z_{k}]
       \vert \Phi_{{\rm GS},B} \rangle}
{2 \langle \Phi_{{\rm GS},B} \vert S^z_{-k} S^z_{k}
\vert \Phi_{{\rm GS},B} \rangle } }\nonumber\\
&=&
\frac{\displaystyle{ 2J(1-\gamma_{k})\langle \Phi_{{\rm GS},B} \vert
\sum_{\langle i,j \rangle \in \Lambda}(-S^x_i S^x_j - S^y_i S^y_j )
\vert \Phi_{{\rm GS},B} \rangle
+ B\langle \Phi_{{\rm GS},B} \vert {\cal M}_\Lambda
\vert \Phi_{{\rm GS},B} \rangle}}
{2N \langle \Phi_{{\rm GS},B} \vert S^z_{-k} S^z_{k}
\vert \Phi_{{\rm GS},B} \rangle }.\mbox{\hspace{1cm}}
\end{eqnarray}
The excitation energy in the thermodynamic limit is given by
\begin{equation}\label{trial:e_energy}
\varepsilon(\mbox{\boldmath $k$}) =
\frac{ 2Jd(\rho_x+\rho_y)(1-\gamma_{k})}
{2 S_\bot^z (\mbox{\boldmath $k$})},
\end{equation}
where $S_\bot^z (\mbox{\boldmath $k$})$ denotes
the structure factor of the N\'eel-ordered ground state,
\begin{equation}
S_\bot^z (\mbox{\boldmath $k$})=
\lim_{B\downarrow 0} \lim_{\Lambda\uparrow \infty}
\langle \Phi_{{\rm GS},B} \vert S^z_{-k} S^z_{k}
\vert \Phi_{{\rm GS},B} \rangle,
\end{equation}
which describes correlations of spins transverse to the magnetic field.
To bound the structure factor for the Heisenberg model from below,
we use the correlation inequality by Shastry,\upcite{Shastry}
\begin{equation}\label{ineq:Shastry}
2 S_\bot^z (\mbox{\boldmath $k$})\ge
\frac{{m_{\rm s}}^2 \sqrt{1-\gamma_k}}{\sqrt{\rho_x+\rho_z}
\sqrt{ 1+\gamma_k }}.
\end{equation}
This bound can be extended to the anisotropic model (\ref{Hamiltonian_XXZ}) in
the form
\begin{equation}\label{ineq:Shastry_XXZ}
2 S_\bot^z (\mbox{\boldmath $k$})\ge
\frac{{m_{\rm s}}^2 \sqrt{1-\gamma_k}}
{\sqrt{\rho_x( 1+\Delta\gamma_k )+\rho_z ( \Delta+\gamma_k)}}.
\end{equation}
Using eqs. (\ref{trial:e_energy}) and (\ref{ineq:Shastry_XXZ}),
we obtain eq. (\ref{ineq:Th3}).
\hfill \qed
\bigskip

\bigskip

Consequently, the lowest spin-wave spectrum is bounded
from above by a gapless $k$-linear dispersion relation.
We thus find that the spin-wave velocity $v_{\rm s}$ is bounded from above
in the form
\begin{equation}
v_{\rm s}\le \sqrt{2d(1+\Delta)(\rho_x + \rho_z)}
(\rho_x+\rho_y )J/{m_{\rm s}}^2.
\end{equation}

The expectation value $\varepsilon (\mbox{\boldmath $k$})$ can be bounded from
below as well.
The transverse structure factor $S^z_\bot (\mbox{\boldmath $k$})$ is
bounded from above in the form\upcite{Dyson,Kennedy1988A}
\begin{equation}\label{ineq:DLS_XXZ}
2 S_\bot^z (\mbox{\boldmath $k$})\le
\left[\frac{ (\rho_x +\rho_y)(1-\gamma_k) }
{\Delta ( 1+\gamma_k )} \right]^{1/2}.
\end{equation}
Note that theorem 4.1 of ref.~\cite{Dyson} is applicable to the
Hamiltonian (2). Using this inequality, we obtain
\begin{equation}
\varepsilon (\mbox{\boldmath $k$})\ge
2dJ \sqrt {\Delta (\rho_x +\rho_y)(1-{\gamma_k}^2) }.
\end{equation}
(We remark that this gives only a lower bound for the expectation
value of the single-mode approximation and not for the spin-wave spectrum.)
Both the upper and lower bounds have the $k$-linear
dispersion relation for small $k$.
We thus find that $\varepsilon (\mbox{\boldmath $k$})$ has the
gapless $k$-linear dispersion relation.

Finally, we discuss the large-$S$ limit of the spectrum. For the Heisenberg
antiferromagnet, we have
\begin{eqnarray}
\lim_{S\rightarrow\infty} m_{\rm s}/S& =&1,\\
\lim_{S\rightarrow\infty} \rho_x/ S^2&=&1,\\
\lim_{S\rightarrow\infty} \rho_y/ S^2&=&
\lim_{S\rightarrow\infty} \rho_z/ S^2=0.
\end{eqnarray}
We show a proof in the Appendix. From these rigorous results, we find that
in the large-$S$ limit the upper bound for the Heisenberg model
coincides with the lower bound and we obtain
\begin{equation}
\varepsilon (\mbox{\boldmath $k$})=2dSJ\sqrt{1-{\gamma_k}^2}.
\end{equation}
This is the same result as that given by the spin-wave
theory.\upcite{Anderson,RKubo}
We thus find that the single-mode approximation gives a precise spin-wave
spectrum in the large-$S$ limit.
For the large-$S$ $XXZ$ model ($0\le \Delta <1$), it is expected
that the ground state becomes the
N\'eel state and we have $\rho_x \simeq S^2$,
$\rho_y \simeq O(S)$, $\rho_z \simeq O(S)$ and $m_{\rm s} \simeq S$, though
we have not proven it. Then we expect that the upper bound
for the spectrum behaves as $2dJS\sqrt{(1+\Delta \gamma_k) (1-\gamma_k)}$.
This spectrum coincides with that given by the linear
spin-wave theory for the $XXZ$ model.\upcite{Nishimori1985}

\bigskip

The author would like to thank Professor M.~Suzuki for helpful discussions
and for critically reading this manuscript.
The author is also grateful to Dr.~T.~Koma and Professor~H.~Tasaki for
useful comments.

\section*{\large Appendix. Some rigorous results for the large-$S$
Heisenberg antiferromagnet}
\hspace*{\parindent}
In this appendix we show bounds for $m_{\rm s}$, $\rho_x$, $\rho_y$ and
$\rho_z$ in the Heisenberg antiferromagnet on square and cubic lattices.
{}From these bounds, we obtain the $S$ dependence of these values
in the $S\rightarrow \infty$ limit.

The quantities $m_{\rm s}$, $\rho_x$, $\rho_y$ and $\rho_z$ are
bounded from above by the norms of local operators. Hence, we obtain
\begin{equation}\label{ineq:m_upper}
m_{\rm s}\le S
\end{equation}
and
\begin{equation}\label{ineq:rho_upper}
\rho_\alpha \le S^2 \mbox{\hspace{1cm}}(\alpha=x,y,z).
\end{equation}
The value $(-\rho_x-\rho_y-\rho_z)$ is the ground-state energy per bond,
whose bounds were shown by Anderson\upcite{Anderson111} in the form
\begin{equation}\label{ineq:GS_bounds}
S^2 \le (\rho_x+\rho_y+\rho_z)\le S(S+1/2d).
\end{equation}

To bound $m_s$ from below, we use Koma and Tasaki's
inequality\upcite{Koma1993A}
\begin{equation}\label{ineq:KomaT}
m_{\rm s} \ge \left( 3 \lim_{\Lambda\uparrow Z^d} {1\over N^2}
\langle \Phi_{{\rm GS},B=0}| {{\cal M}_{\Lambda}}^2
|\Phi_{{\rm GS},B=0}\rangle\right)^{1/2}
\end{equation}
and the inequality by Dyson, Lieb and Simon\upcite{Dyson} and Jord\~ao Nevez
and Fernando Perez\upcite{Jordao}
\begin{eqnarray}\label{ineq:KLS}
\lefteqn{\lim_{\Lambda\uparrow Z^d} {1\over N^2}
\langle \Phi_{{\rm GS},B=0}| {{\cal M}_{\Lambda}}^2
|\Phi_{{\rm GS},B=0}\rangle}\\
& &\ge S(S+1)/3-(\rho_x + \rho_y +\rho_z )^{1/2}
I_d/\sqrt6,\nonumber
\end{eqnarray}
where
\begin{equation}
I_d=\int {dk^d \over (2\pi)^d}
\left( { 1-\gamma_k \over 1+\gamma_k } \right)^{1/2}.
\end{equation}
The integral $I_d$ is finite for $d\ge 2$.
It was shown that the r.h.s. of the inequality (\ref{ineq:KLS}) is positive
for $S\ge1$ in the
two- and three-dimensional systems.\upcite{Jordao,Affleck}
Using eqs.~(\ref{ineq:GS_bounds})-(\ref{ineq:KLS}),
we obtain bounds of $m_{\rm s}$ in the form
\begin{eqnarray}
\lefteqn{\left[S(S+1)-\sqrt{3S(S+1/2d)}I_d /\sqrt2 \right]^{1/2}}\nonumber\\
& &\mbox{\hspace{3.5cm}} \le m_{\rm s} \le S \mbox{\hspace{1.2cm}}
\end{eqnarray}
for $S\ge1$.
Thus we find
\begin{equation}\label{limit:m}
\lim_{S\rightarrow \infty} m_{\rm s}/S=1.
\end{equation}

To bound $\rho_x$, $\rho_y$ and $\rho_z$ from below, we use
eqs.~(\ref{ineq:Shastry}) and (\ref{ineq:DLS_XXZ}),
from which we find
\begin{equation}
{ {m_{\rm s}}^2 \over (\rho_x+\rho_z)^{1/2} }
\le
(\rho_x+\rho_y)^{1/2}.
\end{equation}
{}From symmetry of the Hamiltonian, i.e., $\rho_y=\rho_z$,
we obtain
\begin{equation}\label{ineq:rho_lower}
{m_{\rm s}}^2 \le \rho_x+\rho_y .
\end{equation}
Using eqs.~(\ref{ineq:rho_lower}) and (\ref{ineq:GS_bounds}), we obtain
\begin{equation}\label{ineq:rho_upper2}
{m_{\rm s}}^2 +\rho_y \le S(S+1/2d),
\end{equation}
where we have used $\rho_y=\rho_z$.
Using eqs.~(\ref{ineq:rho_upper}), (\ref{ineq:rho_lower}) and
(\ref{ineq:rho_upper2}),
we obtain bounds for $\rho_\alpha$ ($\alpha=x,y,z$) in the form
\begin{equation}\label{ineq:rho_bounds}
2{m_{\rm s}}^2 - S(S+1/2d) \le {m_{\rm s}}^2 -\rho_y \le \rho_x \le S^2 .
\end{equation}
Using eqs.~(\ref{ineq:rho_bounds}) and (\ref{limit:m}), we find
\begin{equation}
\lim_{S\rightarrow \infty} \rho_x/ S^2 =1
\end{equation}
and
\begin{equation}
\lim_{S\rightarrow \infty} \rho_y/ S^2
=\lim_{S\rightarrow \infty} \rho_z/ S^2 =0.
\end{equation}

\end{document}